\documentclass[twocolumn,prl,showpacs,amsfonts,superscriptaddress,floatfix,nofootinbib,10pt]{revtex4-1}
\usepackage{amsmath}
\usepackage{graphicx}
\usepackage{footmisc}
\pdfoutput=1
\usepackage{color}

\begin{document}
\title {The quantum Hall effect at a weak magnetic field}
\author{Igor N. Karnaukhov}
\affiliation{G.V. Kurdyumov Institute for Metal Physics, 36 Vernadsky Boulevard, 03142 Kiev, Ukraine}
\begin{abstract}
Using a weak limit for the hopping integral in one direction in the Hofstadter model, we show that the fermion states in the gaps of the spectrum are determined within the Kitaev chain. The proposed approach allows us to study the behavior of Chern insulators (CI) in different classes of symmetry. We consider the Hofstadter model on the square and honeycomb  lattices in the case of rational and irrational magnetic fluxes $\phi$, and discuss the behavior of the Hall conductance at a weak magnetic field in a sample of finite size. We show that in the semiclassical limit at the center of the fermion spectrum, the Bloch states of fermions turn into chiral Majorana fermion liquid when the magnetic scale $ \frac{1}{ \phi} $ is equal to the sample size N. We are talking about the dielectric-metal phase transition, which is determined by the behavior of the Landau levels in 2D fermion systems in a transverse magnetic field. When a magnetic scale, which determines the wave function of fermions,  exceeds the size of the sample, a jump in the longitudinal conductance occurs. The wave function describes non-localized states of fermions, the sample becomes a conductor, the system changes from the dielectric state to the metallic one.  It is shown, that at $1/\phi>$N the quantum Hall effect and the Landau levels are not realized, which makes possibility to study the  behavior of CI in irrational magnetic fluxes.
\end{abstract}
\maketitle

\section{Introduction}

The Harper-Azbel`-Hofstadter Hamiltonian \cite{Har,Az,Hof} gives possibility to investigate the behavior of the Chern systems both in rational and irrational magnetic fluxes through the unit cell $\phi$. In the Hofstadter model in the semiclassical limit, the edges of the spectrum  have the structure of the Landau levels \cite{a}. The quantum Hall effect is  a clear demonstration  of topological states in the physics of condensed matter \cite{Ex0,Ex1,Ex2}. The quantum Hall effect is realized in semiclassical limit, with a large  magnetic scale $q$, that determines splitting of a Bloch band into a fine structure of the spectrum with $q$-topological subbands  ($q \sim 10^4$ for the case of experimentally realizable magnetic fields). The value of $q$ inverse proportional to a magnetic field, therefore larger $q$ correspond to weak magnetic fields. In \cite{AD} the quantum Hall effect is studied numerically in the presence of correlated disorders at weak magnetic field. How the energy levels in 2D lattice fermion systems evolve with the magnetic field is one of the tasks that is solved in this paper.

A non-trivial topology of the ground state provides the quantization of the Hall conductance, which can be interpreted as the Chern number.  In the case of a rational magnetic flux, a topological state of the Harper-Hofstadter Hamiltonian is characterized by the Chern numbers and chiral edge gapless modes, realized the CI state \cite{H,1,K1,K2}. Filled $r$-bands with the Chern numbers $C_\alpha$ yield a Hall conductance $\sigma_{xy}= (e^2 /h)C_r, C_r= \sum_{\alpha=1}^r C_\alpha$. In the case of an irrational magnetic flux, the Bloch states of the fermions and the Brillouin zone are not defined and the Chern number as the topological order of the 2D fermion system is also not defined. It difficult to calculate the topological numbers using the system-dependent approach based on successive rational approximants of the irrational flux \cite{D1,D2}. The spectral structures of the Fibonacci quasicrystals can be used to study the Bloch states and the spectrum of fermions in a magnetic field, and the gaps in these structures can be labeled with topological numbers. The state of the system in an irrational flux is result of its evaluation using the rational approximation of the magnetic flux by the Fibonacci quasicrystals \cite{F1,F2}.

We study topological properties of the energy spectra of the 1D quasiperiodic systems, describing also Bloch fermions in magnetic field. At a rational flux, when the regime of strong periodic potential is realized, a topological trivial band splits into magnetic subbands. In semiclassical limit these subbands form the Landau levels.
The aim of this article is to answer the question of how irrational magnetic fluxes manifest themselves in the quantum Hall effect. We show that the behavior of the fermion system is determined by its topology. It turns out that the transition from rational to irrational fluxes, when the magnetic scale $q$ is larger than the linear size of the sample N, leads to a rearrangement of the fermion spectrum. At $q>$N the Landau levels transform to dispersion ones, insulator state is not realized, at $q \gg$N free fermion states are realized. In the semiclassical limit, irrational fluxes "kill" the quantum Hall effect, so for a given sample (with its characteristic size), there is a minimal value of a magnetic field, which limits the realization of the quantum Hall effect.

\section{The model}

We will analyze a model of 2D CI in a transverse magnetic field determined in the framework of the Harper-Azbel`-Hofstadter Hamiltonian \cite{Har,Az,Hof}
\begin{eqnarray}
&&{\cal H}= \sum_{x-links}\sum_{j} t^x(j) a^\dagger_{j}  a_{j+1}+ \sum_{y-links}\sum_{j} t^y(j) a^\dagger_{j} a_{j+1}+\nonumber \\
&& \sum_{z-links}\sum_{j} t^z(j) a^\dagger_{j} a_{j+1} + H.c.
    \label{eq:H}
\end{eqnarray}
where $a^\dagger_{j}$ and $a_{j}$ are the Fermi operators determined on a site $j=\{x_j,y_j\}$. The Hamiltonian (\ref{eq:H}) takes into account the hoppings of spinless fermions located at the nearest -neighbor lattice sites with different hopping amplitudes $t^x (j)=t$ (along the x-links), $t^y (j)=\exp[2 \pi i(x_j-1)\phi]$ (along the y-links) in a square lattice, $t^x (j)=t$, $t^y (j)=(-1)^{x_j} \exp[2 \pi i(x_j-1)\phi]$  in an alternative square lattice, and $t^x (j)=t $, $t^y (j)=t^z(j)=\exp[i \pi (x_j-1)\phi]$ in a honeycomb lattice with x,y,z-links. A magnetic flux through a unit cell $\phi = \frac{H }{ \Phi_0}$ is determined in the quantum flux unit ${\Phi_0=h/e}$, a homogeneous magnetic field $H$ is defined by the vector potential $\textbf{A} = H x \textbf{e}$. The $x$-links determine coupling between $y$-chains in the square lattices and the zig-zag $\xi$-chains in a honeycomb lattice, the vector potential is directed along $\textbf{e}_y$ or $\textbf{e}_\xi=\{ \frac{1}{2}, \frac{\sqrt3}{2}\}$ directions. We consider the 2D fermion system in the stripe geometry with open boundary conditions for the boundaries along the y-direction in the square lattices and along the zig-zag chains in a honeycomb lattice. For such geometry the 2D model is reduced to the 1D system (along the x-direction of length N).

\section{Square lattices. Topological structure of the spectrum }
\subsection{A square lattice. \\A rational flux $\phi =\frac{p}{q}$}
   \begin{figure}[tp]
     \centering{\leavevmode}
\begin{minipage}[h]{.9\linewidth}
\center{
\includegraphics[width=\linewidth]{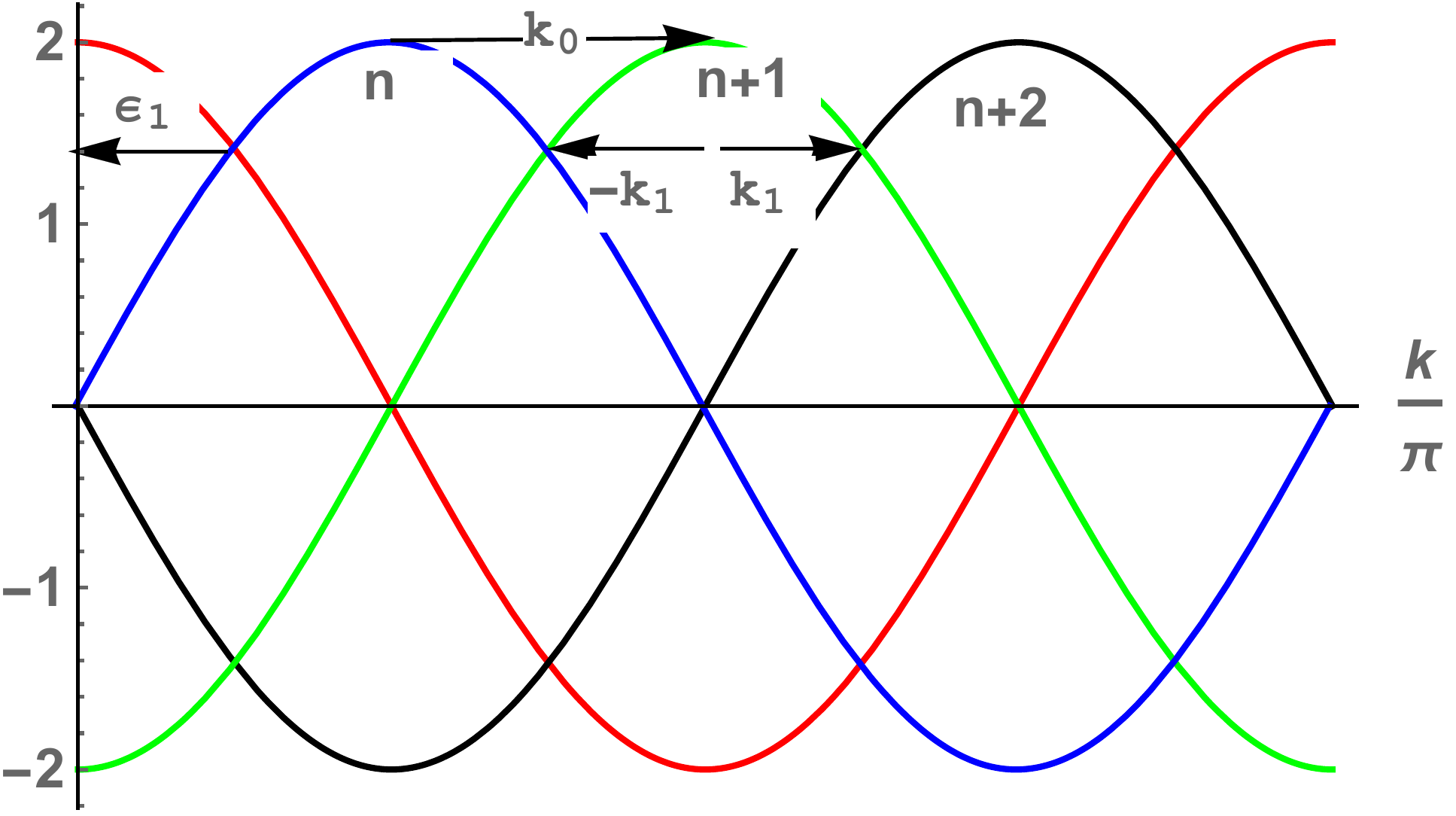} a)\\
                  }
    \end{minipage}
     \centering{\leavevmode}
\begin{minipage}[h]{.9\linewidth}
\center{
\includegraphics[width=\linewidth]{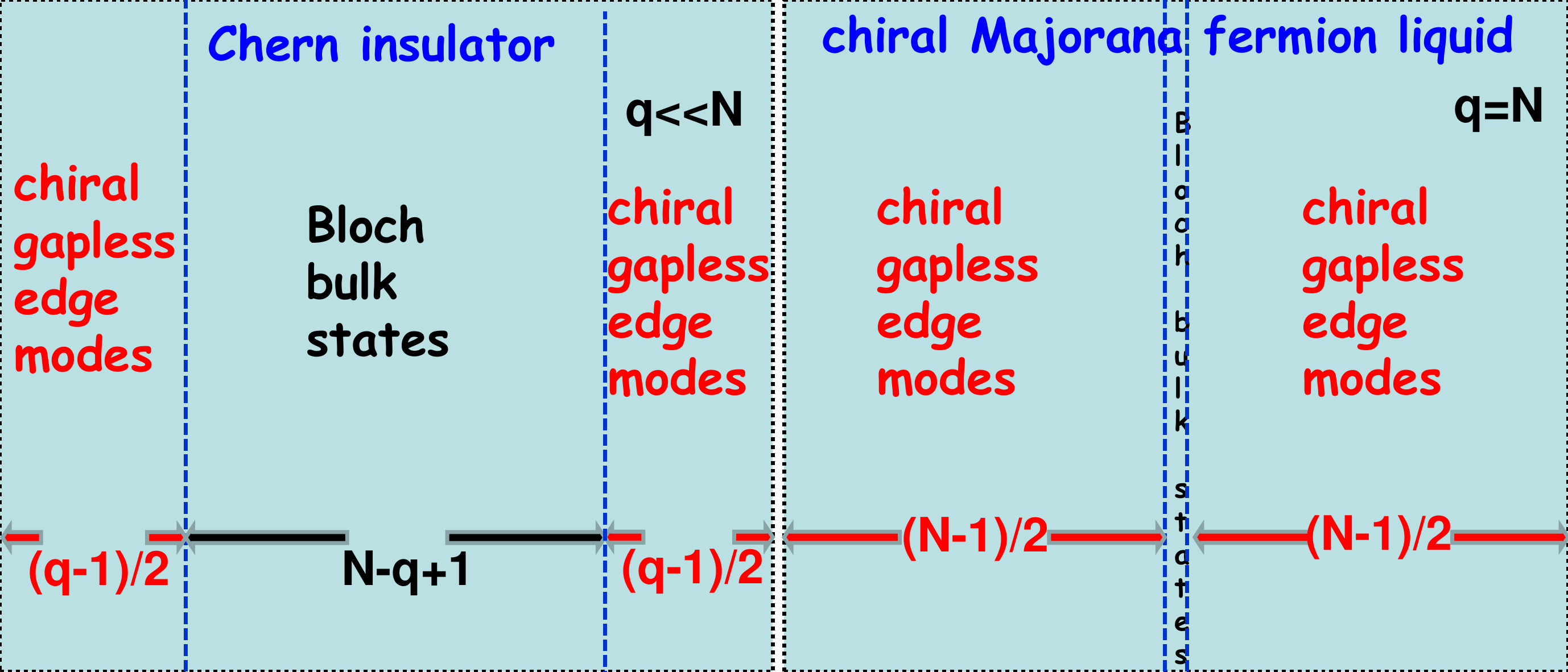} b)\\
                  }
    \end{minipage}
\caption{(Color online)
The spectra of $ n, n+1, n +2 $ chains at $ t = 0 $ a),  the energies of fermions in them are shifted by $ k_0=\frac{2\pi}{q}$, they, marked with color lines, intersect at the points $n k_0 \pm k_1, \epsilon_1 $. Formation of the chiral Majorana fermion liquid as a result of domination of chiral edge states b), here the magnetic flux $\phi=\frac{1}{q}$ with odd $q$, $N$ is the linear size of the sample.
  }
\label{fig:1}
\end{figure}
\begin{figure}[tp]
     \centering{\leavevmode}
     \begin{minipage}[h]{1.\linewidth}
\center{
\includegraphics[width=\linewidth]{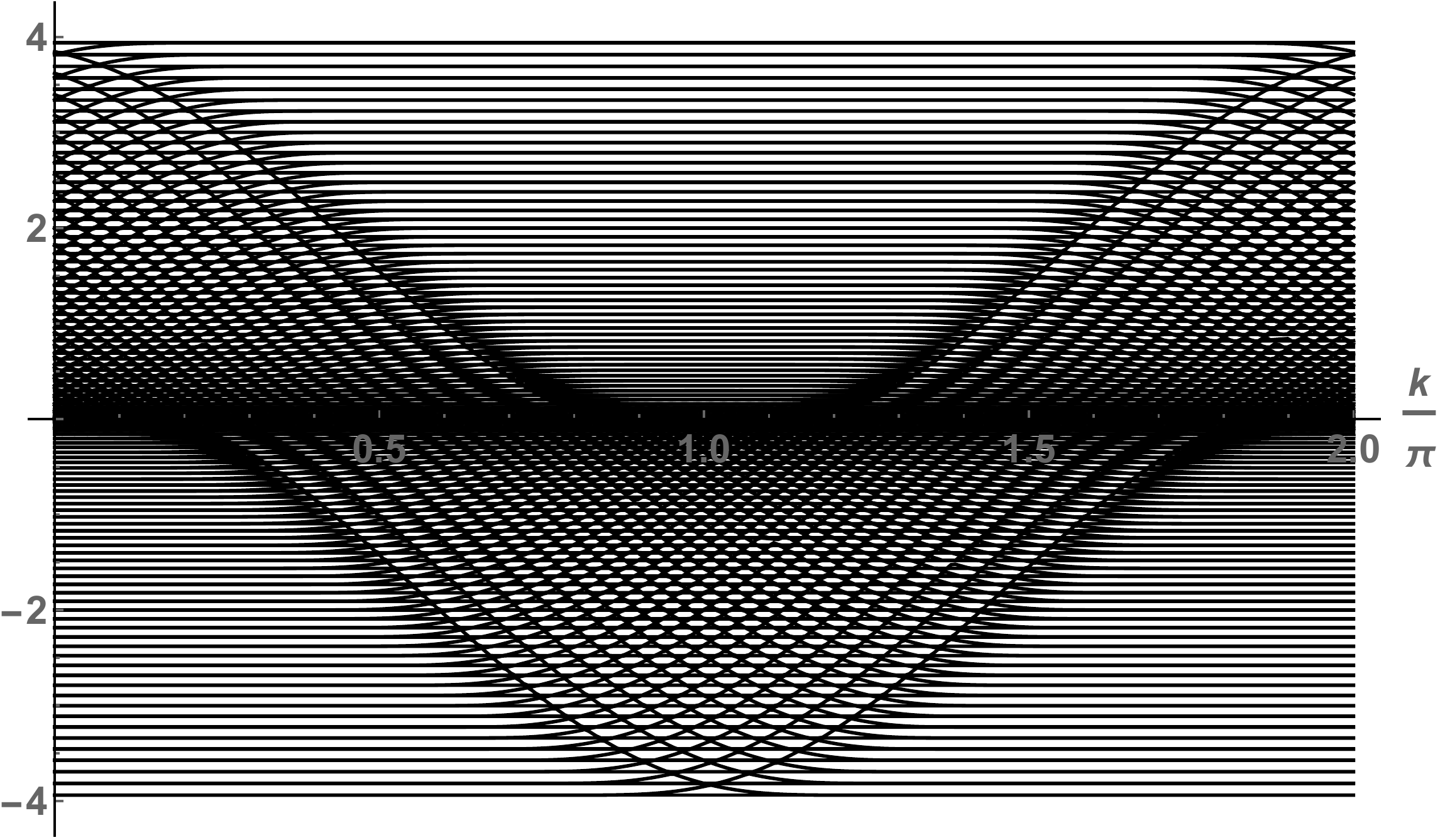} a)\\
                  }
    \end{minipage}
\begin{minipage}[h]{1\linewidth}
\center{
\includegraphics[width=\linewidth]{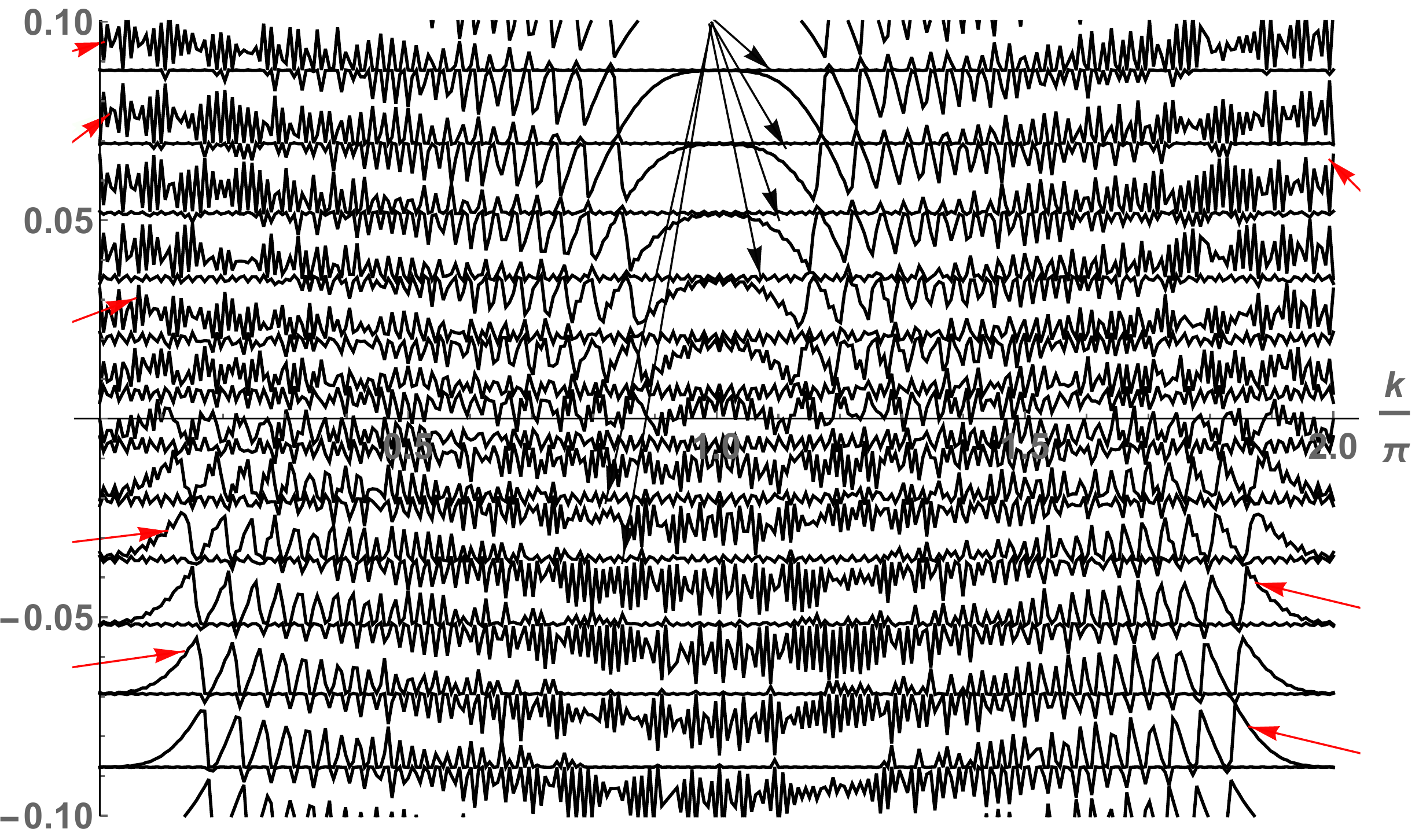} b)\\
                  }
    \end{minipage}
     \centering{\leavevmode}
\begin{minipage}[h]{1\linewidth}
\center{
\includegraphics[width=\linewidth]{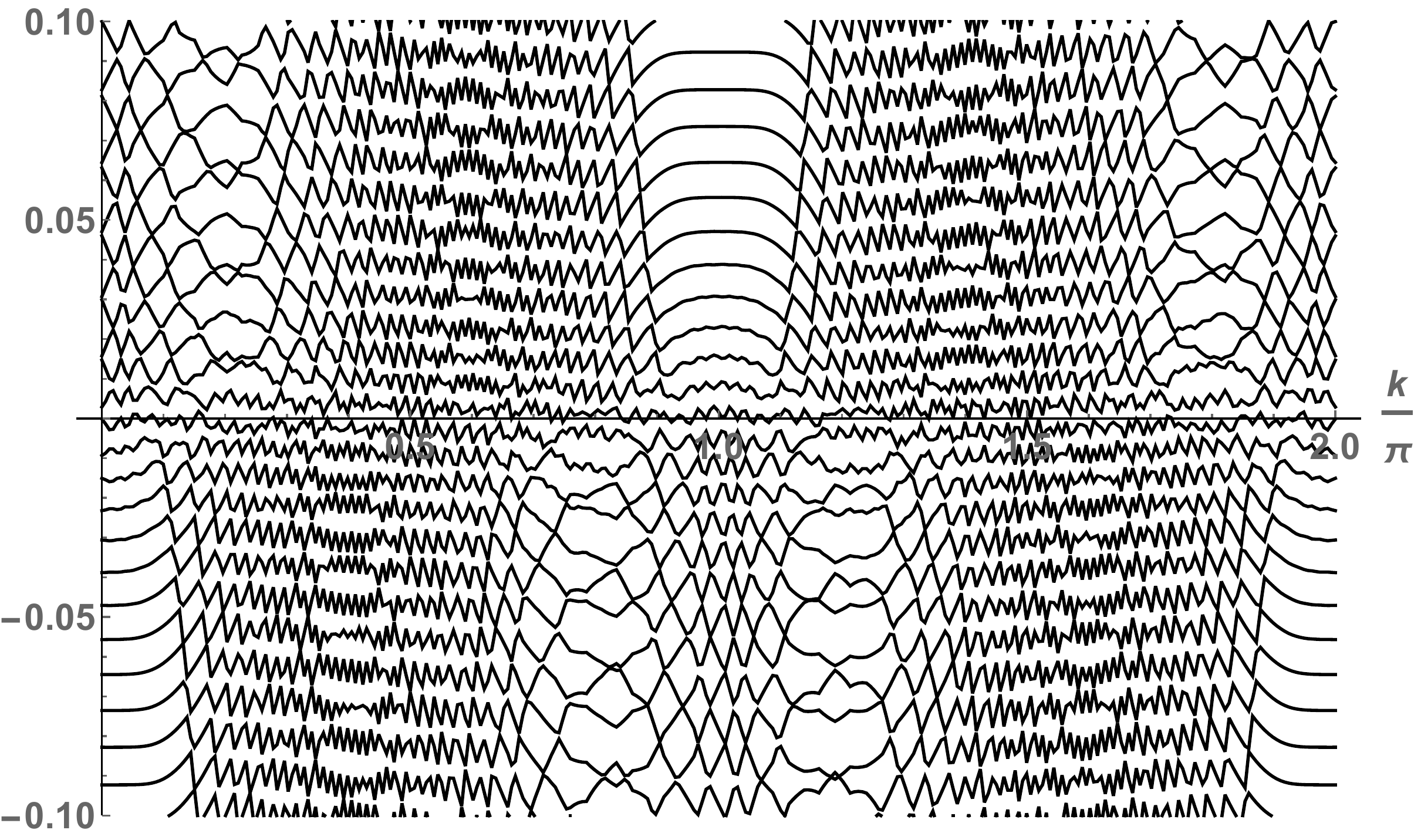} c)\\
                  }
    \end{minipage}
\caption{
The fermion spectrum containing Landau-levels and chiral edge modes  calculated on a square lattice at $t=1$ and N=400 and different $q$:  $q=100$ a), and low energy spectra at $q=200$ b) and q=400 c). The Bloch (the quasi-Landau levels) and chiral Majorana fermion states are marked by black and red arrows, respectively b), the spectrum corresponding to the chiral Majorana fermion liquid c).
}
\label{fig:2}
\end{figure}

Let us consider the $q$-subbans, splitting from one Bloch band for a rational flux $\phi= \frac{p}{q}$, here $p$ and $q$ are coprime integers. At $q>2$ a magnetic field breaks a time reversal symmetry,  the CI state is realized \cite{H,1,K1,K2}.  In this case  a topological trivial Bloch band splits into $q$-magnetic subbands  with non-trivial topological index $C_\alpha$ ($\alpha$ is the index of a subband in a fine structure of the spectrum).
Two systems that correspond to weak ($t\to 0$) and strong ($t=1$) coupling limits of the Hamiltonian (\ref{eq:H}) with an arbitrary magnetic scale can be continuously deformed into each other, the topological numbers during deformation will not change. The topological numbers of subbands in the fine structure of the spectrum depend on the value of the magnetic flux $\phi$ and in a wide range of values do not depend on $t$. Numerical calculations shown that the Chern numbers do not depend on the bandwidths forming the fermion spectrum when the hopping integral $t$ varies over a wide range \cite{21,IK}. The excitation spectrum of a sample with open boundary conditions also includes chiral modes localized at the boundaries.

We focus our attention on the weak coupling limit, fermions propagate in the $y$-chains, weak tunneling between them. The fermions of the nearest chains with the energies shifted by $ 2\pi \phi$ tunnel, the gaps open at the intersection points of the energies. The gaps in the spectrum are determined by the values of the effective tunnel integral for fermions of different chains with Fermi energy (the Fermi energy lies in the gap in CI).

We will take into account the fermion states in the gaps \cite{IK,IKa} with the energies equal $\epsilon_s = \pm 2\cos k_s$  in the $t\to 0$ limit for $k_s$, here $k$ is the y-projection of the wave vector, $k_s = s \pi \phi$, $s =1,...q$. At these energies, the states of bulk fermions are not realized.
From numerical calculations, it follows that the gap is determined by the distance between the chains $\delta$  and the tunneling constant t $\Delta (\epsilon) = \zeta t^\delta +0(t^{\delta +1})$, here $ \zeta $ decreases from 2 to 1 with increasing $q$.

Note that only two states of fermions in each $y-n$-chain which have momenta $k_n=k_0 (n-1) \pm k_s$ and energies $\epsilon_s$ (where $k_0=\frac{2\pi}{q}$), tunnel between the chains (see in  Fig.\ref{fig:1}a)). These states are determined by the Fermi operators $a^\dagger (k_n)\Rightarrow a^\dagger_n, a(k_n)\Rightarrow a_n$, here $n$ numerates the $y$-chain along the $x$-direction (see in Fig.\ref{fig:1}a)). Tunneling of fermions can be described by the low-energy Hamiltonian ${\cal H}_{eff}= \tau(\delta) \sum_{n=1}^{N-\delta} a^\dagger_n a_{n+\delta}+ H.c.$, where $\tau(\delta)$ is the effective hopping integral  between fermions located at the sites on the distance $\delta$, $n$ defines a lattice site along the $x$-direction or the number of the $y$-chain. Using the operators
$\chi_{n}=a_{n}+ a_{n}^\dagger$ and $\gamma_{n}=\frac{ a_{n}- a_{n}^\dagger}{i}$, we redefine the Hamiltonian as $i \frac{\tau (\delta)}{2} \sum_{n=1}^{N-\delta}( \chi_{n} \gamma_{n+\delta} -\gamma_{n} \chi_{n+\delta})$.
The Majorana operators $\gamma_n$ are defined by a Clifford algebra $\{\gamma_n,\gamma_m \}=2 \delta_{n,m}$ and $\gamma_n=\gamma^\dagger_n$.

We must take into account that the hoppings of fermions between chains are realized only between particles with different chirality.  The low energy effective Hamiltonian, which determines excitations of Majorana fermions in the gap with the energy $\varepsilon$, has the following form {\cite{IK,IKa,0}
\begin{eqnarray}
&&{\cal H}(\epsilon) = \frac{i}{2}\tau (\delta) \sum_{n=1}^{N-\delta}\gamma_{n} \chi_{n+\delta},\nonumber \\
&&{\cal H}(-\epsilon) = \frac{i}{2}\tau (\delta) \sum_{n=1}^{N-\delta}\gamma_{n+\delta} \chi_{n}.
\label{eq:H2}
\end{eqnarray}

The total Hamiltonian ${\cal H}(\epsilon)+{\cal H}(-\epsilon)$ is the $U(1)$-symmetric. The Hamiltonian (\ref{eq:H2}) describes the chain of isolated dimers of pairing Majorana fermions with the energy $\pm \frac{\tau (\delta)}{2}$ and zero energy Majorana fermions at the boundaries \cite{Ki1}. The gap in fermion spectrum is equal to $\tau (\delta)$. $2\delta$ edge modes with different velocities are localized at the boundaries, they determine the Hall conductance and the Chern number $C(\epsilon_F)=\pm\delta$ (here $\epsilon_F$ is the Fermi energy).
At $\epsilon=\to 0$ the Hamiltonian ${\cal H}(\epsilon)+{\cal H}(-\epsilon)$(\ref{eq:H2}) determines gapless fermion states in the center of the spectrum. According to bulk edge correspondence the Chern number  $C_r$ equal to the total number of edges modes localized at a boundary, when the Fermi energy lies in the r-th gap from the bottom of the band (r numerates filled subbands).

In the semiclassical limit $p=1$, $q\to \infty$, that corresponds to experimentally realizable magnetic field, we obtain the value of q for different H, such for a lattice constant $a=3{\AA}$  $q=4.6\cdot10^8/H(gauss)$ or $q=4.6\cdot 10^6$ for $H=10^2$ gauss and $q=4.6\cdot 10^5$ for $H=10^3$ gauss. At $t=1$ the edges of the spectrum are characterized by the Landau levels $\epsilon_n=4 \pi \phi (n+1/2)$ (see in Fig.\ref{fig:2}a)). In the $t\to 0$ limit the fermion spectrum is discrete with the gaps at $\epsilon_n = (n\pi\phi)^2$, here $n$ numerates the energy levels,
 $|C_n|=n$ and $\sigma_{xy}=\frac{e^2}{h}n$.
The Chern number has a maximum value $C_{max}(\epsilon)=\frac{q-1}{2}$ (for odd $q$) or $C_{max}(\epsilon)=\frac{q}{2}-1$ (for even $q$) near the center of the spectrum at $\epsilon\to 0$. The edge modes localized at the boundaries occupy a region of size sample (q-1) or (q-2) (see in Fig.\ref{fig:1}b)), the Bloch states of fermions occupy the rest region of the sample N-q+1 or N-q+2. At q=N the Bloch states of fermions are not realized, and a phase transition to the state of the chiral Majorana fermion liquid occurs. The chiral Majorana fermion liquid states corresponds to abnormal large value of  $C_{max}(\epsilon \to 0)$ at $q$=N.

We have illustrated said above in Figs \ref{fig:2}b),c), where calculations of the low energy spectrum are shown at  N=400, $q= 200$  and $q= 400$. At $q=200$ the Bloch states (that determine the Landau levels in the semiclassical limit) co-exist with the chiral Majorana fermion states, see in Fig. \ref{fig:2}b),  at $q=400$ c) only chiral Majorana states are realized. The Majorana states are quasi-one-dimensional, they localized along the x-direction and phase separated, see in Fig. \ref{fig:2}c). This phase transition is characterized by certain conductance properties in 2D samples i.e. the vanishing of the longitudinal conductance $\sigma_{xx}=0$ and the bulk Hall conductance $\sigma_{xy}\simeq\frac{Ne^2}{2h}$. The state of a chiral Majorana fermion liquid describes the behavior of spinless (non-interacting) fermions in the framework of the Hofstadter model with $n \simeq \frac{1}{2}$ and $q\simeq N$. The case $q>$N actually corresponds to the realization of an irrational magnetic flux.

\subsection{An irrational flux}
\begin{figure}[tp]
     \centering{\leavevmode}
\begin{minipage}[h]{.49\linewidth}
\center{
\includegraphics[width=\linewidth]{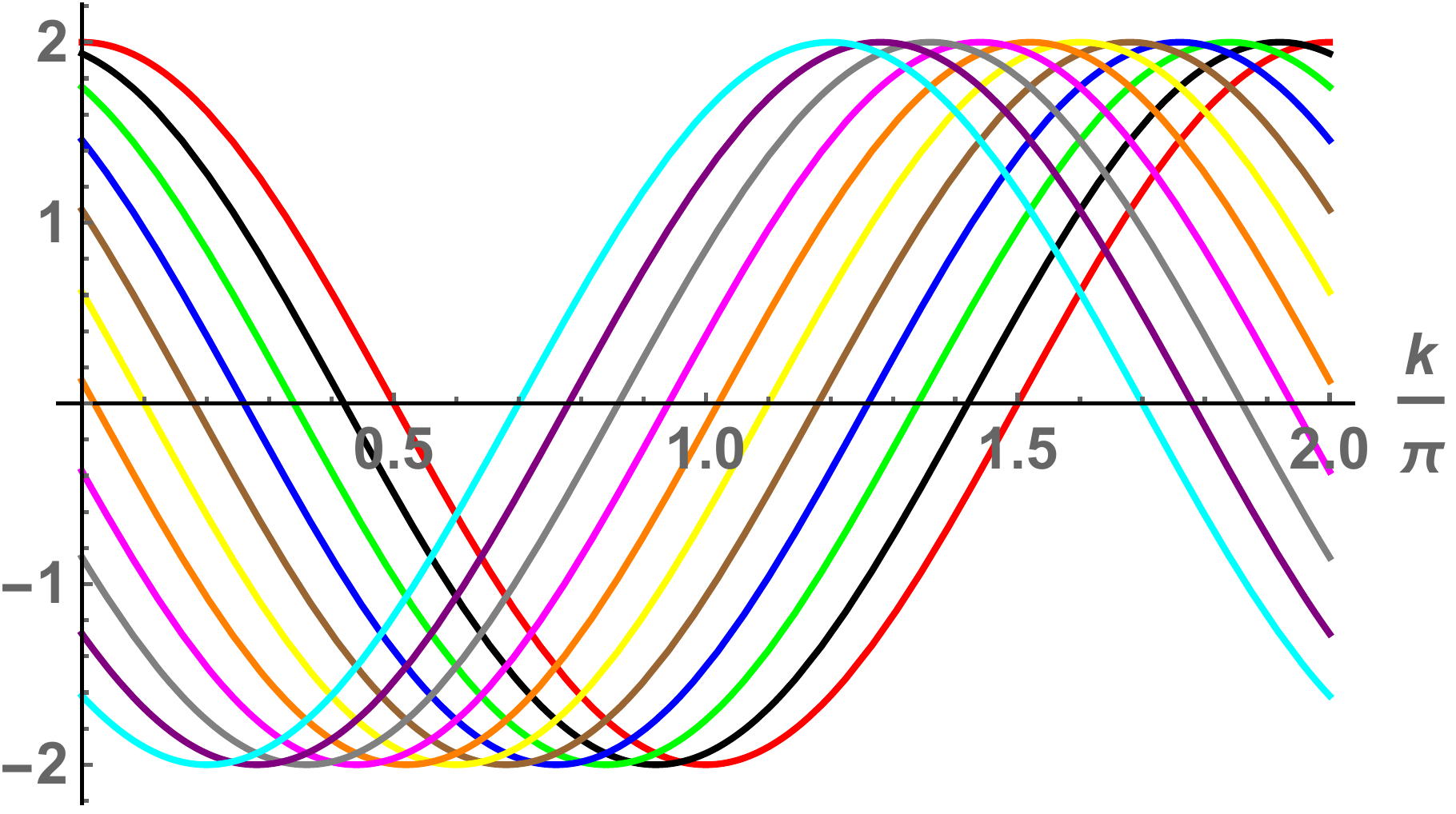} a)\\
                  }
    \end{minipage}
\caption{(Color online)
The spectrum of fermions for N=11 and $q=25$ at $ t = 0 $, the energies of fermions in the chains marked with color lines.
  }
\label{fig:3}
\end{figure}
\begin{figure}[tp]
     \centering{\leavevmode}
\begin{minipage}[h]{,48\linewidth}
\center{
\includegraphics[width=\linewidth]{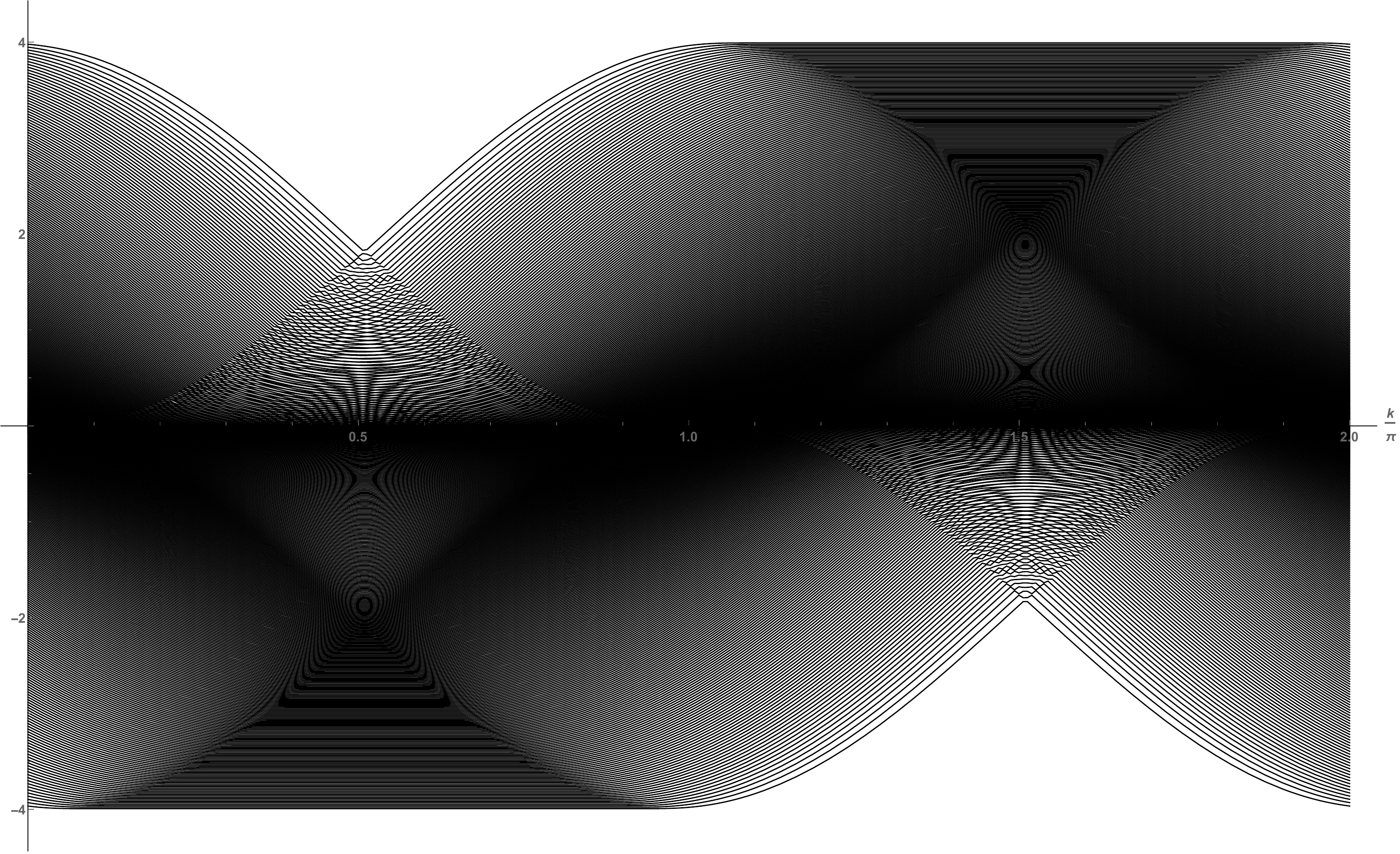} a)\\
                  }
    \end{minipage}
    \begin{minipage}[h]{.48\linewidth}
\center{
\includegraphics[width=\linewidth]{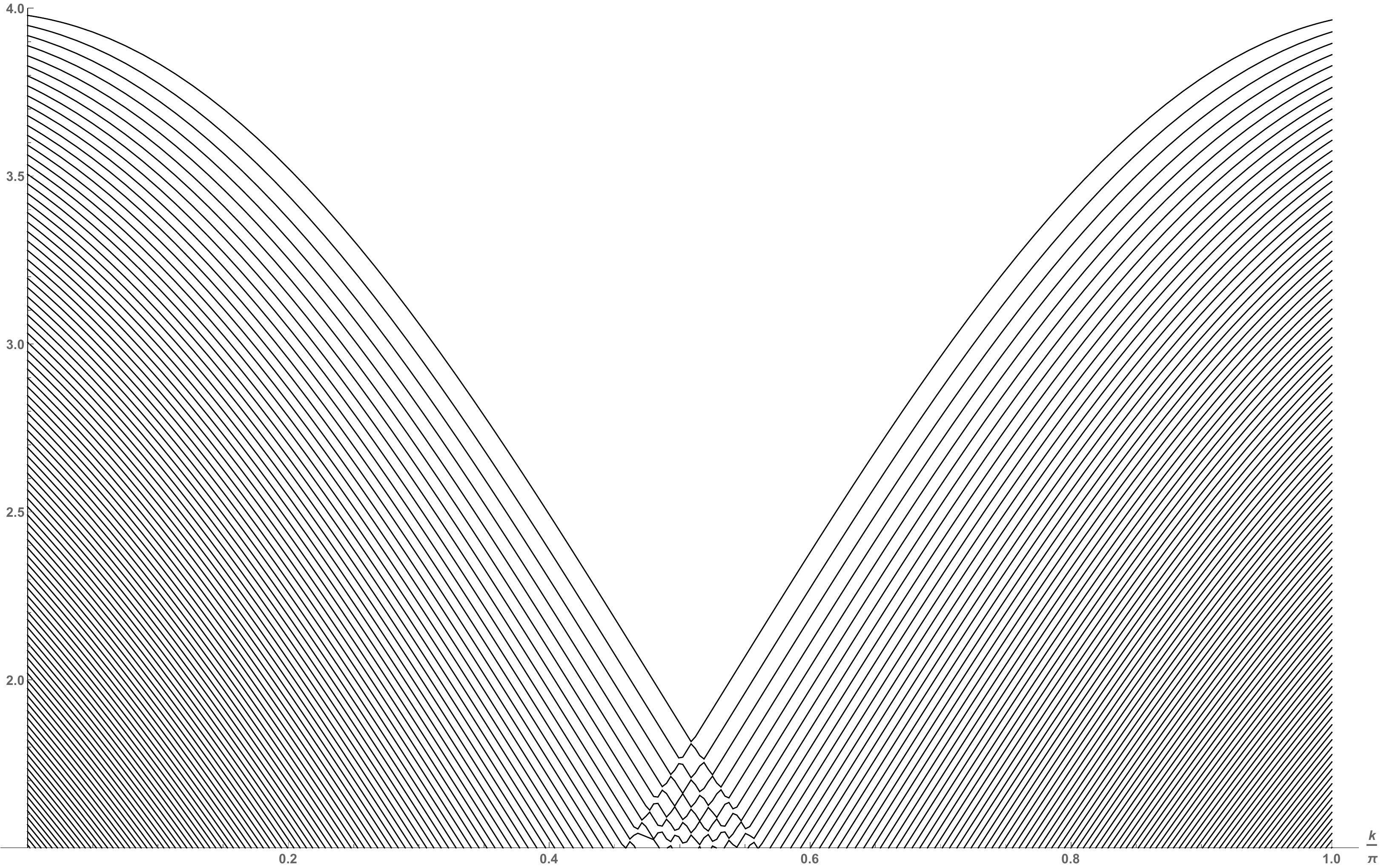} b)\\
                  }
    \end{minipage}
     \centering{\leavevmode}
    \begin{minipage}[h]{\linewidth}
\center{
\includegraphics[width=\linewidth]{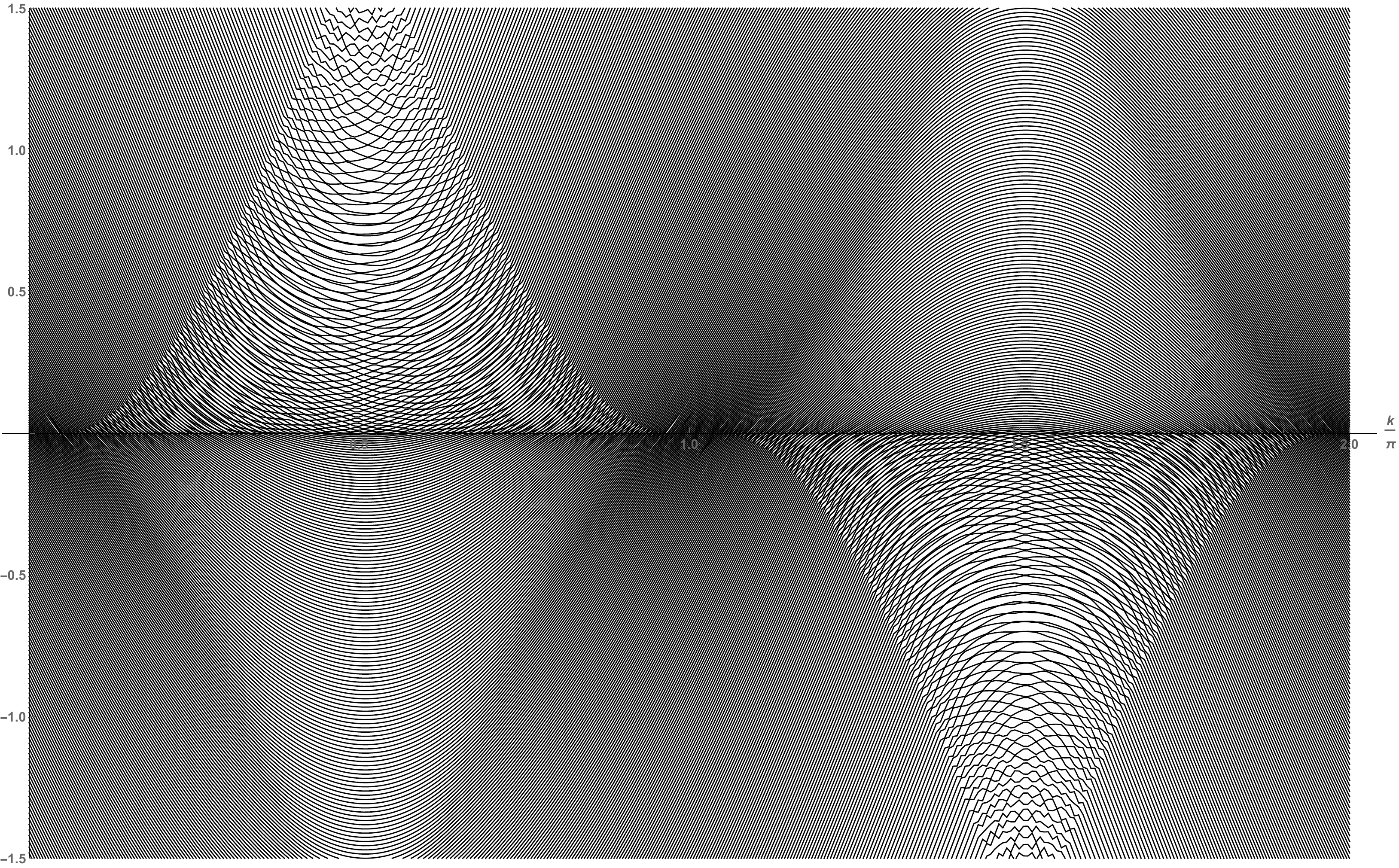} c)\\
                  }
    \end{minipage}
\caption{
The spectrum of fermions calculated on a square lattice at $t=1$, N=400  $q=813$ a), a high-energy spectrum with quasi-Landau levels near the upper edge of the spectrum for $k/\pi=[0,1]$ b), a low energy spectrum c).}
\label{fig:4}
\end{figure}

We consider an evolution of the chiral Majorana fermion liquid state in the Hofstadter model with an irrational flux. In this case, the chiral gapless edge modes determine the topological properties of both the spectrum and the Hall conductance.
The magnetic scale $q$, defined by $\phi$, must be greater than N, in this case the irrational flux is determined within  an accuracy of $\frac{1}{N}$. The transition from rational to irrational magnetic fluxes is determined by the ratio $\kappa =\frac {q}{N}$; $\kappa <1$, $\kappa >1$  correspond to rational, irrational fluxes, and $\kappa = 1$  corresponds to the transition from one to another. The point $ \kappa = 1 $ can be considered as a phase transition point, since the symmetry of the spectrum changes. As can be seen from Figs \ref{fig:2},\ref{fig:4}, at an irrational magnetic flux the inversion of the fermion spectrum (depending on $k_y$) is  broken.  For $ H \to 0 $ $\kappa \to \infty$; therefore, for $ \kappa = [1, \infty] $, the states near a center of the spectrum  evolve from the chiral Majorana fermion liquid state to free fermions.

For illustration of forming of the fermion spectrum for $q>$N we consider the case $t=0$ (see in Fig.\ref{fig:3}). At $q>$2(N-1), $\delta$-crossing point corresponds to N-$\delta$ intersections of the energies of fermions in the chains separated at distance $\delta$, $\delta $=N-1 for the lowest energy intersection. The gaps open only at the edges of the spectrum, forming dispersive Landau levels (see in Figs\ref{fig:4}a),b) and compare with Fig.\ref{fig:2}a)). In contrast to a rational flux, the insulator phase is not realized due to the dispersion of the levels. In the center of the spectrum, the gaps are not open, only the singularities of the fermion states occur with wave vectors corresponding to energy intersections.
Such the lowest energy intersection is realized at one value of the wave vector (see in Fig.\ref{fig:4}a)). At $q>10$N the region of gap spectrum near the edges of the spectrum is small, the fermion states are the Bloch states with the spectrum of free fermions. The Landau levels near the edges of the spectrum at $ q \leq $N  become the dispersion levels for $ q>$ N, while the low-energy states, such as the chiral Majorana fluid, are transformed into free fermion states at $ q \gg $N.

\subsection{An alternative square lattice.\\A rational flux}

\begin{figure}[tp]
    \centering{\leavevmode}
\begin{minipage}[h]{1\linewidth}
\center{
\includegraphics[width=\linewidth]{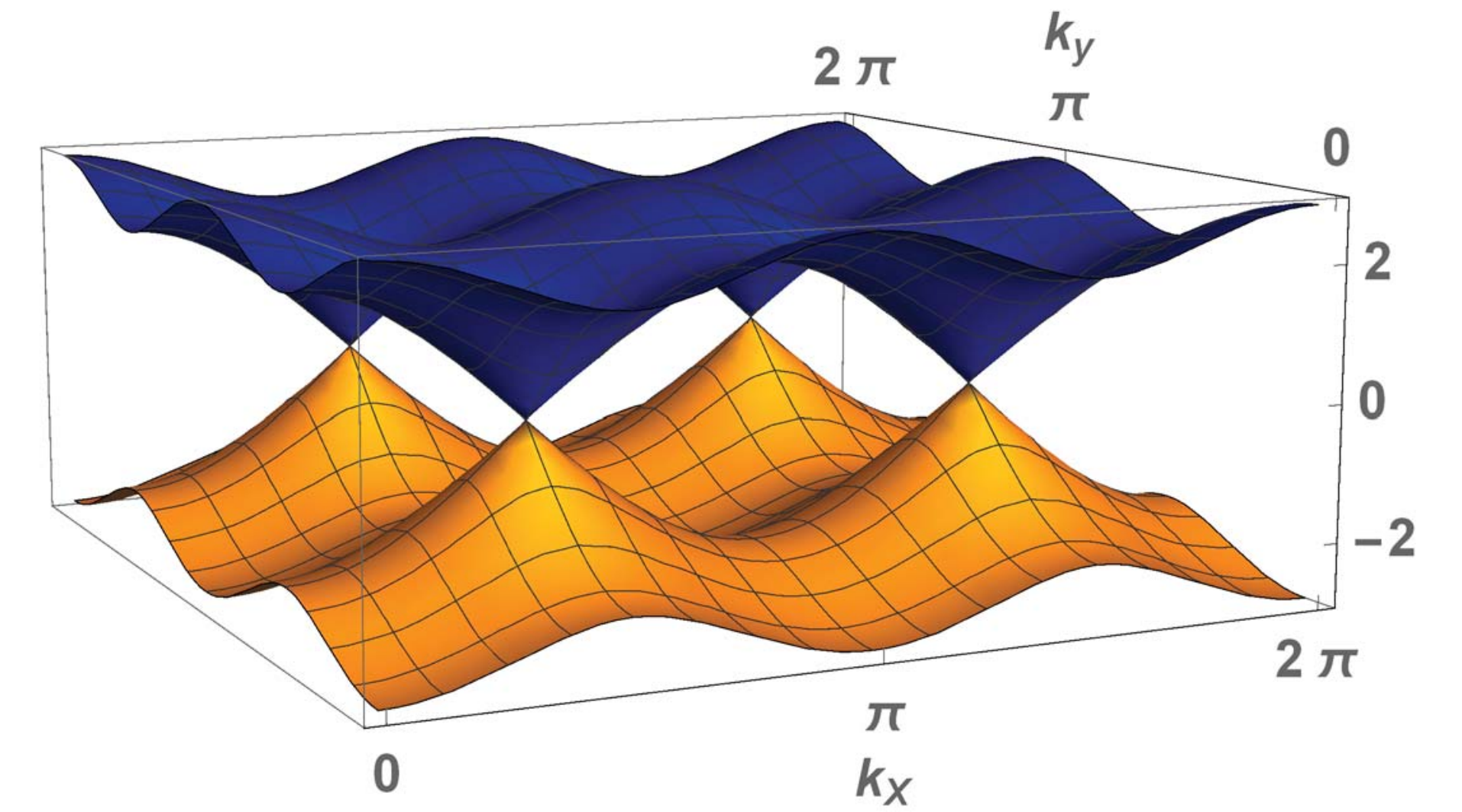} \\ a)
                  }
\end{minipage}
\begin{minipage}[h]{1\linewidth}
\center{
\includegraphics[width=\linewidth]{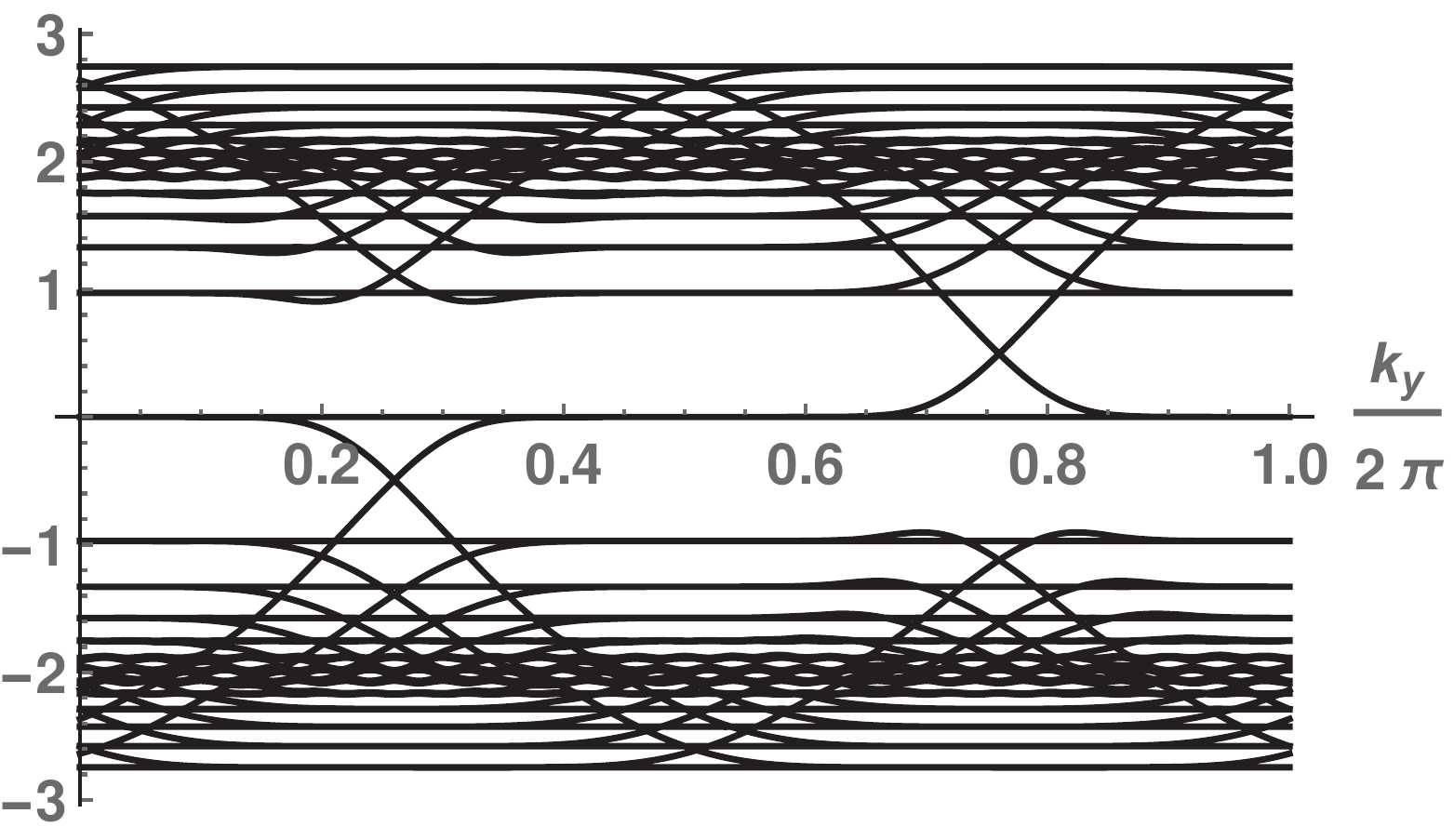} \\ b)
                  }
\end{minipage}
\caption{(Color online)
The spectrum of fermions that corresponds to the ground state with a double cell:
with periodic boundary conditions (boundary along y-direction) and H=0, four Dirac points at $\{\pm \frac{\pi}{2}, \pm \frac{\pi}{2},\}$ a); with open boundary conditions at a rational magnetic flux N=300, $q=50$, the structure of the quasi-Landau, Dirac levels for a rational magnetic flux with chiral gapless edge modes b).
       }
\label{fig:5}
\end{figure}

The fermion spectrum of an alternative square lattice $\epsilon(\textbf{k}) = \pm 2 \sqrt {\cos^2 k_x + \cos^2 k_y}$  describes a gapless state with four Dirac points at $\{\pm \frac{\pi}{2},\pm\frac{\pi}{2}\}$ (see in Fig. \ref {fig:5}a)). In the semiclassical approximation with rational values of $q$, the spectrum of spinless fermions contains levels and gapless edge modes (see in  Fig. \ref{fig:5}b)).  At half-filling the gapless state is realized for arbitrary rational $q$.  The spectrum is topological symmetric with respect to zero energy, since $|C (\epsilon_F)| =|C(-\epsilon_F)|$.

The spectrum of spinless fermions around the edges of the band has the structure of the Landau levels (see in Fig.\ref{fig:5}b)). At $t=1$ the energy levels are approximated by the following form $\epsilon_n = \pm \sqrt{2}\left [2-2\pi\phi (n+\frac{1}{2})\right]$,  with the corresponding coefficient at order $2\pi \phi$ $\omega =\sqrt{2}$ ($\omega =\frac{1}{\sqrt{3}}$ in a honeycomb lattice \cite{a}, for comparison). In the $t \to 0$  limit the energy of the gaps are equal to $\epsilon_n =2-( 2\pi n \phi)^2$. In the semiclassical limit the modulus of the Chern number for n-filled subbands near the edges of the spectrum (n counts the subbands from the edges of the band) is equal to $|C_n| = 2n$. The energies of the next-nearest chains are shifted by $\frac{4\pi}{q}$, so the Chern numbers of are determined with scale 2n.
The energies of the levels near the zero energy or half-filling have another energy scale. At $t=1$ the spectrum is approximated as $\epsilon_n = \pm 4\sqrt{\pi\phi (n+\frac{1}{2})}$ with doubled $\omega$ (the coefficient at order $\sqrt{2\pi\phi}$), $\omega =3^{1/4}$  in a honeycomb lattice \cite{a}. In the $t\to 0$ limit the gaps in the spectrum open at $\epsilon_n= 2\pi\phi (2n+1)$, where n numbers the subbands from the center of the spectrum. The modulus of the Chern number for the n-filled or empty subbands (for the fermion states near the center of the band) is equal to $ |C_n|= 2n+1$. This behavior of the Chern number is characteristic of the Dirac states.

\section{A honeycomb lattice}
\begin{figure}[tp]
     \centering{\leavevmode}
\begin{minipage}[h]{1\linewidth}
\center{
\includegraphics[width=\linewidth]{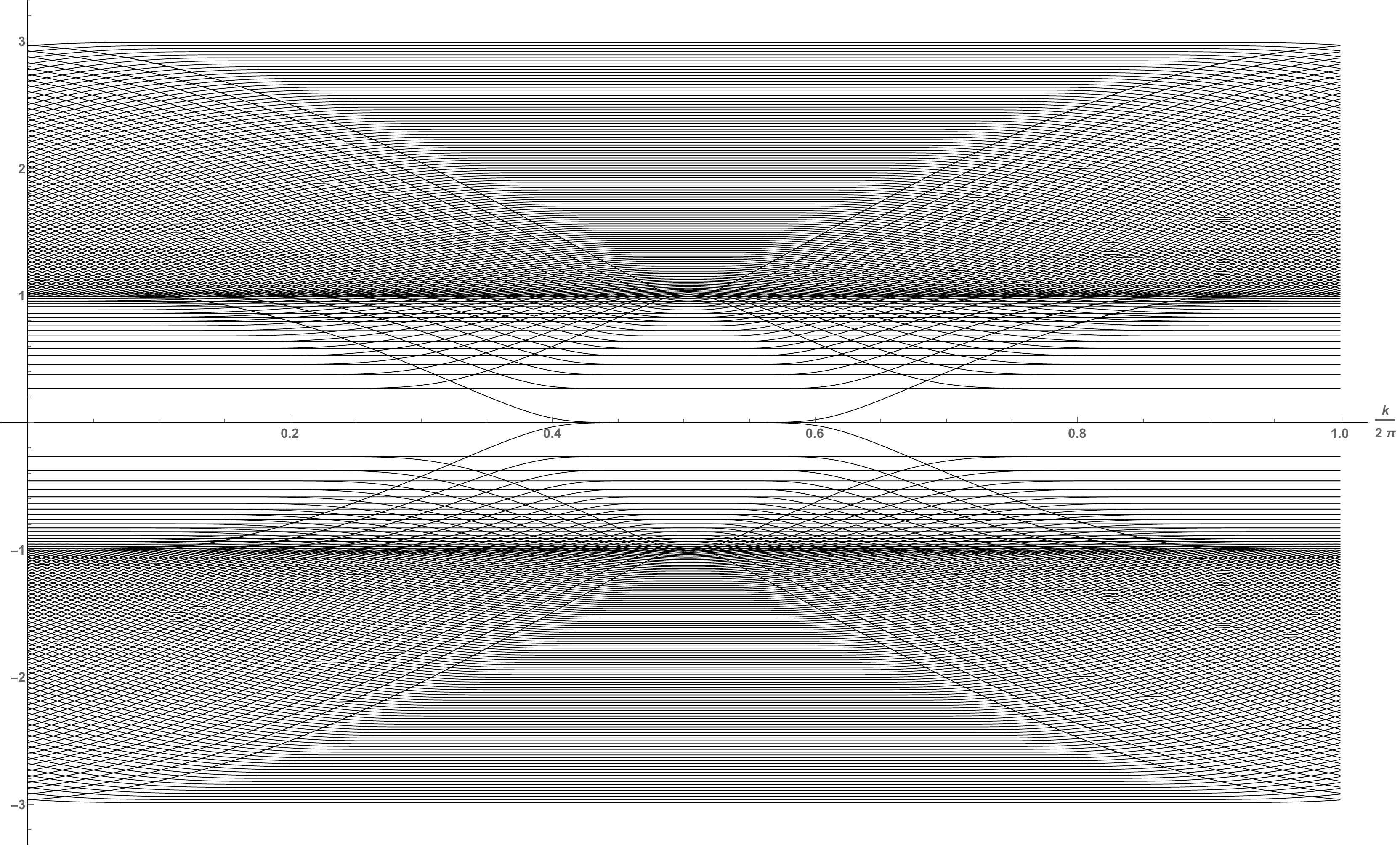} a)\\
                  }
    \end{minipage}
    \begin{minipage}[h]{1\linewidth}
\center{
\includegraphics[width=\linewidth]{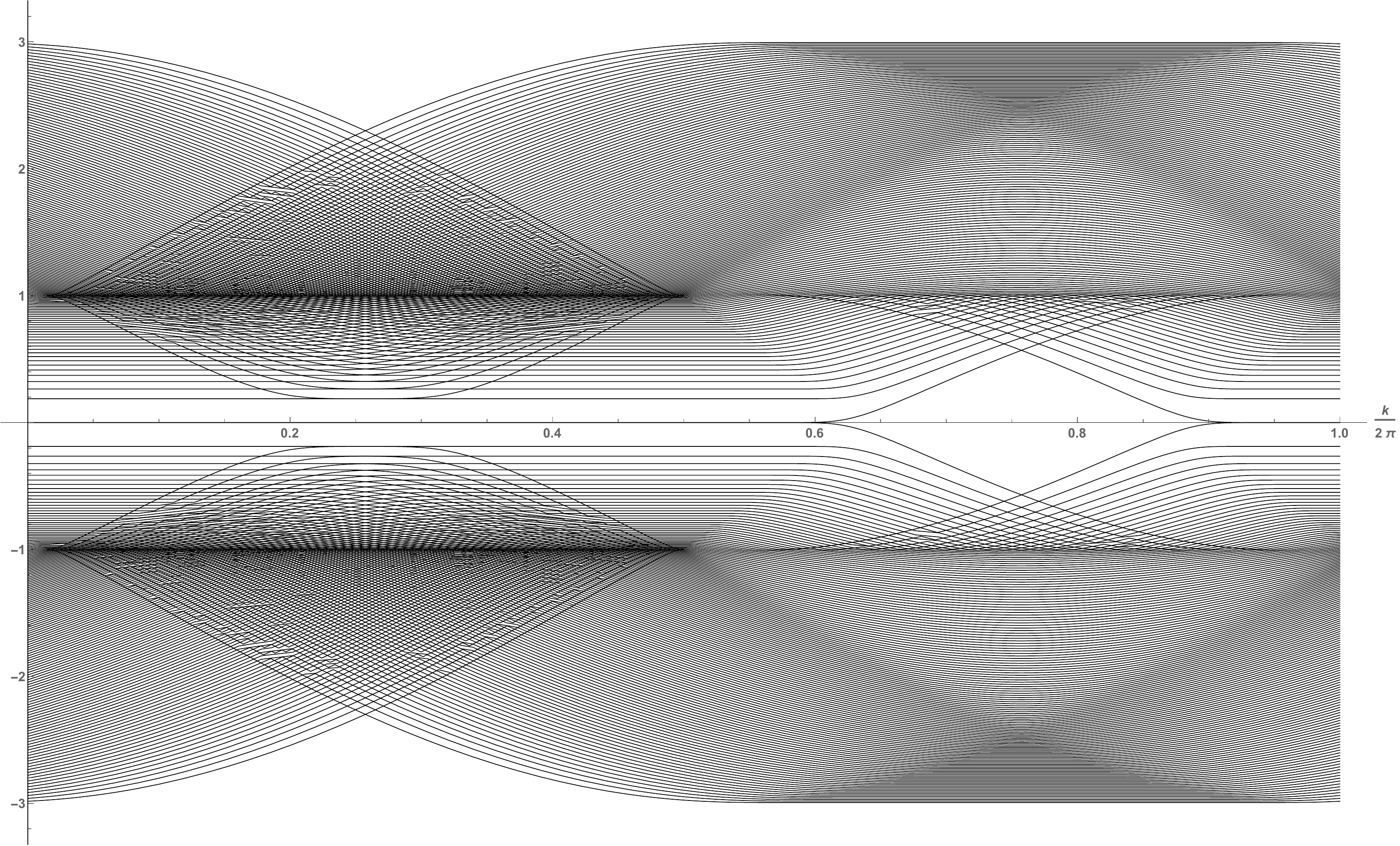} b)\\
                  }
    \end{minipage}
     \centering{\leavevmode}
\caption{
The spectrum of fermions calculated on a honeycomb lattice at $t=1$, N=150; $q=150$, $\kappa=1$ a) and $q=307$, $\kappa =2.047$ b).}
\label{fig:6}
\end{figure}

The spectrum of graphene is characterized by two Dirac points, which determine the behavior of graphene.
In graphene the distance between the Landau levels is very large compared to the Zeeman splitting \cite{23}, so  in the Hamiltonian (\ref{eq:H}) this term is not taken into account. In this case the Hamiltonian (\ref{eq:H}) describes degenerated states of electrons with different spin in an external magnetic field, the model reduces to spinless fermions.
In the Kitaev model \cite{Ki2}, the low energy excitations of which are described in the frame of the tight-binding model (\ref{eq:H}) with anisotropic hopping integrals at $H=0$, the ground state is unstable in gapless phase at half-filling. The Zeeman splitting opens the gap in the spectrum, stabilizes the topological state \cite{Ki2}.  In graphene (with isotropic hopping integrals in (\ref{eq:H})) the ground state is also unstable in the weak magnetic field limit, an arbitrary small magnetic field opens the gap in the center of the spectrum \cite{IK1}. The topological state with zero energy Majorana fermion states and zero Chern number is realized in  magnetic field at half-filling.

The topological structure of the spectrum can be calculated in a weak coupling limit with respect to $t$ in a manner analogous to that of a square lattice. In this case, the edge modes are described by the Hamiltonian (\ref{eq:H2}) and the Chern number is determined by the corresponding value of $\delta$. For a rational flux $\phi$ in the $t \to 0$ limit, the energies of fermions in the chains are shifted by $\pi \phi$ and intersect at the points $k_s= \pi s\phi$, $\varepsilon_s=\pm 2 \cos (\frac{\pi s \phi}{2})$, $\varepsilon_s=\pm 2 \sin (\frac{\pi s \phi}{2})$, here $s=1,...,q$. The set $k_s$, $\varepsilon_s \neq 0$  determines the gaps in the fermion spectrum for given rational flux $\phi$. In semiclassical limit the maximal values of $\delta$ and the Chern number, equal to $\frac{q}{2}$, are realized at quoter (or three quarters) filling. With these fillings in the case $q=$N the chiral Majorana fermion liquid is realized on a honeycomb lattice (see in Fig.\ref{fig:6}a)). The Landau and Dirac levels are dispersionless ones.
Numerical calculations of the fermion spectrum are shown in Fig.\ref{fig:5}, for arbitrary magnetic flux, the spectrum is a mirror symmetric (including to edge states).
At point $\kappa =1$ the inversion symmetry of the spectrum is broken, the Landau and Dirac levels transform to dispersion levels in the case of an irrational magnetic flux (see in Fig.\ref{fig:6}b) at  $\kappa =2.047$). The insulator state is not realized and, as result, the quantum Hall effect disappears.

\section{Conclusion }

In the frame of conclusion, let us consider the experimental realization of irrational magnetic fluxes in the Chern system. The sample of a stripe geometry is determined by the width of the stripe N$a$. The magnetic scale $q$ is  inversely proportional to the amplitude of the magnetic field $H$, $q a=0.14$cm at $H=10^2$ gauss, $q a=0.014$cm at $H=10^3$ gauss. We have shown that at an irrational magnetic flux with $q>$N, the quantum Hall effect is not realized. The minimum value of the magnetic field at which the quantum Hall effect is realized, is determined by the sample size, namely, $q =$ N. For Na$<$0.14 cm (Na$ <$0.014 cm)  the quantum Hall effect is realized at $H>10^2$ gauss ($H>10^3$ gauss). The measurement of the Hall conductance in the magnetic field in samples of small sizes makes it possible to establish  the  behavior of CI under irrational magnetic fluxes.
Considering square and  honeycomb lattices, we have shown that the nature of the quantum Hall effect does not depend on the crystal symmetry of the 2D system, it is universal in this sense.

The studies were also supported by the National Academy of Sciences of Ukraine within the budget program 6541230-3A "Support for the development of priority areas of scientific research".


\section*{References}
\begin{thebibliography}{31}
\bibitem{Har} P.G.Harper, Proceedings of the Physical Society A,  68, (1955), 874.
\bibitem{Az} M.Ya.Azbel`, JETP, 19, (1964), 634.
\bibitem{Hof} D.R.Hofstadter,  Phys.Rev.B, 14, (1976), 2239.
\bibitem{a}  Y.Hatsuda, {Perturbative/nonperturbative aspects of Bloch electrons in a honeycomb lattice}, (2018), arXiv:1712.04012v2 [hep-th].
\bibitem{Ex0} K.v.Klitzing, G.Dorda, and M.Pepper, Phys.Rev.Lett., 45, (1980), 494.
\bibitem{Ex1} K.S.Novoselov, A.K.Geim, S.V.Morozov, D.Jiang, M.I.Katsnelson, I.V.Grigorieva, S,V.Dubonos and A.A.Firsov, Nature, 438, (2005), 197.
\bibitem{Ex2} Y.Zhang, Y.W.Tan, H.L.Stormer and P.Y.Kim,  Nature, 438, (2005), 201.
\bibitem{AD} D.N.Sheng, Z.Y.Weng and  X.G.Wen, Phys.Rev.B, 64,(2001), 165317 and references therein
\bibitem{H}  F.D.M.Haldane, Phys.Rev.Lett., 61, (1988), 2015.
\bibitem{1}  Y.Hatsugai, J.Phys.:Condens.Matter, 9, (1997), 2507
\bibitem{K1} I.N.Karnaukhov, Physics Letters A, 381, (2017), 1967.
\bibitem{K2} I.N.Karnaukhov, Scientific Reports, 7, (2017), 7008
\bibitem{D1} I.Dana, Phys.Rev.B, 89, (2014), 205111.
\bibitem{D2} G.Amit and I.Dana, Phys.Rev.B, 97, (2018), 075137.
\bibitem{F1} D.Damanik and A.Gorodetski,  Commun. Math. Phys., 305, (2011), 221.
\bibitem{F2} H.Hiramoto and M.Kohmoto, Int.J.Mod.Phys.B, 6, (1992), 281.
\bibitem{21} M.Sato, D.Tobe and M.Kohmoto, Phys.Rev.B, 78, (2008), 235322.
\bibitem{IK} I.N.Karnaukhov, J.Phys.Commun., 1, (2017), 051001.
\bibitem{IKa} I.N.Karnaukhov, Europhysics Letters, 124 (2018) 37002.
\bibitem{0}
Ñ.Malciu, L.Mazza, and C.Mora, \emph{$4\pi$ and $8\pi$ dual Josephson effects induced by symmetry defects}, arXiv:1901.03342v1 [cond-mat.mes-hall] 2019.
\bibitem{Ki1} A.Yu.Kitaev, Phys.Usp., 44, (2001), 131.
\bibitem{23} V.P.Gusynin and S.C.Sharapov, Phys.Rev.Lett., 95, (2005), 146801.
\bibitem{Ki2} A.Yu.Kitaev, Annals of Physics, 321, (2006), 2.
\bibitem{IK1} I.N.Karnaukhov, Phys.Lett.A , 383 (2019)  2114.
\end{thebibliography}
\end{document}